# Sentiment Analysis on YouTube: A Brief Survey

**Muhammad Zubair Asghar[1], Shakeel Ahmad[2], Afsana Marwat[1], Fazal Masud Kundi[1]**

[1]Institute of Computing and Information Technology (ICIT), Gomal University, D. I. Khan, Pakistan.

[2]Faculty of Computing and Information Technology in Rabigh (FCITR), King Abdul Aziz University (KAU) Saudi Arabia.

zubair@gu.edu.pk, shakeel_1965@yahoo.com, afsana.marwat84@gmail.com, paitank@live.com

## Abstract

*Sentiment analysis or opinion mining is the field of study related to analyze opinions, sentiments, evaluations, attitudes, and emotions of users which they express on social media and other online resources. The revolution of social media sites has also attracted the users towards video sharing sites, such as YouTube. The online users express their opinions or sentiments on the videos that they watch on such sites. This paper presents a brief survey of techniques to analyze opinions posted by users about a particular video.*

## 1. Introduction

The popularity of social media is increasing rapidly because it is easy in use and simple to create and share images, video even from those users who are technically unaware of social media.

There are many web platforms that are used to share non-textual content such as videos, images, animations that allow users to add comments for each item [1]. YouTube is probably the most popular of them, with millions of videos uploaded by its users and billions of comments for all of these videos.

In social media especially in YOUTUBE, detection of sentiment polarity is a very challenging task due to some limitations in current sentiment dictionaries. In present dictionaries there are no proper sentiments of terms created by community. It is clear from the studies conducted by [2, 3] that the web traffic is 20% and Internet traffic is 10% of the total YOUTUBE traffic. There are many mechanisms of YOUTUBE for the judgment of opinions and views of users on a video. These mechanisms include voting, rating, favorites, sharing. Analysis of user comments is a source through which useful data may be achieved for many applications. These applications may include comment filtering, personal recommendation and user profiling. Different techniques are adopted for sentiment analysis of user comments and for this purpose sentiment lexicon called SentiWordNet is used [4, 5]. In this paper a brief survey is performed on "sentiment analysis using YOUTUBE" in order to find the polarity of user comments.

The whole paper is organized as follows: In Section-2 Survey Framework of sentiment analysis is discussed. In section 3 Evaluation is described. Section 4 gives conclusion.

## 2. Survey Framework

The Survey framework (Fifure 1) of our paper covers three main issues, namely, event classification, detection of sentiment polarity, and predicting YouTube comments. The detailed discussion about each component of the framework is given in the following section

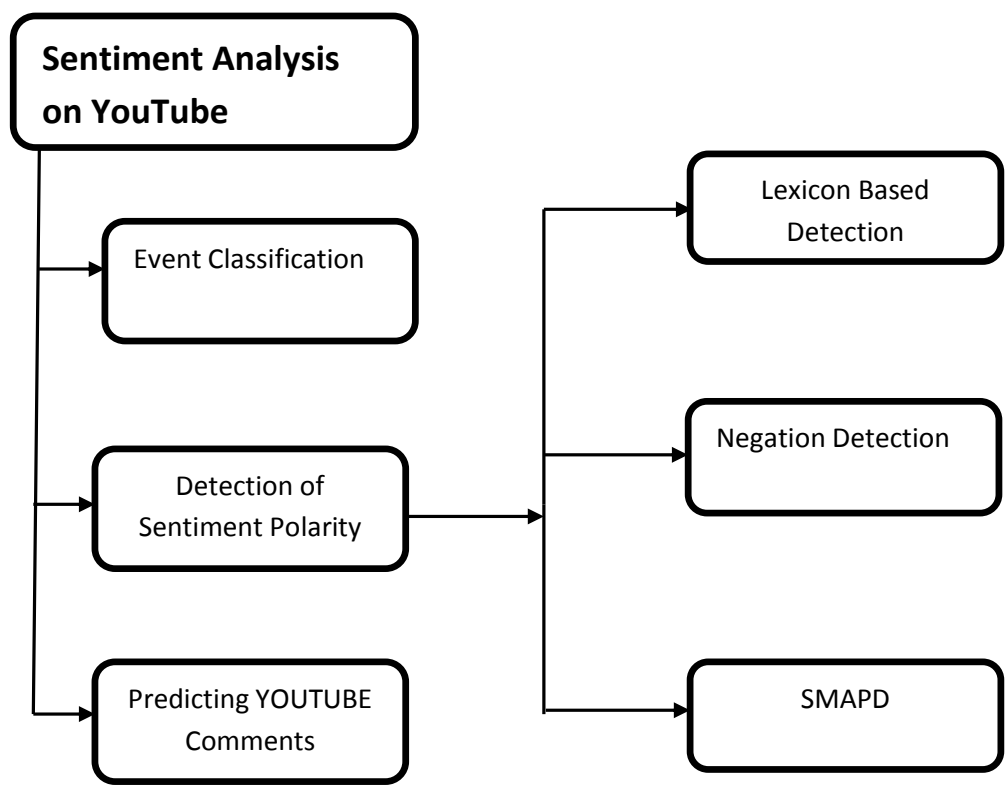

**Fig.1. Sentiment Analysis on YOUTUBE**

## 2.1 Event Classification

To categorize the events as critical, high, medium, low or noise is known as Event Classification. Events are classified as Standard or Special Events.

A technique i.e. a text processing pipeline is used to find out a collection for categories of video event automatically. For this, video titles and description are analyzed and then WordNet is applied so that category may be selected [6].

Users can upload their own videos to YOUTUBE. There is a title and description of each video and this title and description have some information about that video. But this information provided by title and description is not sufficient to understand the video contents [5, 4, 7]. So this information is called weak label information. To make it useful information, first, structure of language used in titles and description must be clear. Secondly a method should be developed to filter categories so that relevant categories may be retained. In order to solve these problems "text processing pipeline" is developed [8, 9] to filter category labels of video events.

## 2.2 Detection of Sentiment Polarity

In this section we address three main issues, namely, lexicon based sentiment analysis, negation detection and Social Media Aware Phrase Detection (SMAPD) pertaining to detect sentiment polarity in YouTube comments.

### 2.2.1 Lexicon based detection

Lexicon oriented techniques for the detection of sentiment polarity are based on different lexicons, such as, WordNet and SentiWordNet (SWN) [10, 11,12].

For the detection of sentiments from user comments, an existing WordNet dictionary and a specific list are used. The specific list contains terms and phrases. These terms and phrases express user opinions[1, 13].The SentiWordNet (SWN) is a document resource which contains a list of English terms which have been attributed a score of positivity and negativity[14].The SWN assigns polarity to each term and phrase of WordNet [12, 15].For analyzing user sentiment in comments, SentiWordNet [6, 9, 16 ] is used.

Through sentiment analysis positive and negative opinions, emotions can be identified. User comments can be analyzed by using SWN and customized social media specific phrase list [1, 17].

When a user enters a comment, the system detects the number of those words having sentiment. A sentiment score is calculated by the number of entries for every word present in the list. If there is a single entry of a word then that word is assigned a polarity with the highest score. If there are more than one entries of a term, the scores of each class are averaged for normalization. For example, there are four scores for a term in a sentence. To find the sentiment polarity of the entire comment, the average score of all terms in the comment is calculated and the term having highest frequency is assigned either positive or negative polarity. Table 1 shows sentiment scores for terms [1, 4, 12, 16].

**Table 1.**User comment. [Adopted from Smitashree Choudhury et al.]

| Comment | Sentiment polarity score (pos., neg., neutral) |
|---|---|
| I think it would benefit religious people to see things like this, not just to learn about our home, the Universe, in a fun and easy way, but also to understand that non- religious explanations don't leave people hopeless and helpless as they think: they inspire people with awe, understanding and a thirst for exploration. Can you ask for more? | **benefit**: 0.0: 0.125: 0.875 |
| | **fun**: 0.0: 0.0: 1.0 |
| | **easy**: 0.0: 0.625: 0.375 |
| | **understand**: 0.375: 0.125: 0.5 |
| | **leave**: 0.0: 0.0: 1.0 |
| | **hopeless**: 0.0: 0.75: 0.25 |
| | **inspire**: 0.0: 0.125: 0.875 |
| | **awe**: 0.5: 0.125: 0.375 |
| | **understanding**: 0.0: 0.0: 1.0 |
| | **thirst**: 0.25: 0.0: 0.75 |
| Document class | Neutral |

### 2.2.2 Negation Detection

Negation is an operation used in grammar. A statement is negated by replacing it with opposite statement [18]. A negative form of statement describes the falsity of the basic statement.

**Negation of Verb Phrases:** Sentences can first be negated through negation of verb phrase [18]. For this negation, negative adverb ***not*** is used. For example; (i) They have books. (Positive) (ii) They do not have books (Negated). **Negation of Noun Phrases**: Sentences can secondly be negated through noun phrase negation [18]. Noun phrases can be negated by using the quantifying determiner ***no*** in front of the noun phrase. For example; (i) They have money. (Positive) (ii)They have no money (Negated).

**Negating Adjective Phrases:** Adjective phrases are negated by using the negative adverb "**not**" in front of the adjective phrase [18]. For example (i)The book is heavy (Simple sentence) (ii) The book is not heavy(Negated).

**Other Quantifiers used for Negation**: Sentences can also be converted into negative form by using different quantifiers, such as, no one, nobody, and none, never, nowhere, nothing, hardly, a scarcely [18]. The lexicon consisting of negation phrases having a list of high-frequency terms in a negative context is maintained and then phrase level polarity score calculation is performed in such a way that if the input is identified as positive phrase then previously determined sentiment score is recovered from negative to positive [1, 16, 19].

### 2.2.3 Social Media Aware Phrase Detection (SMAPD)

In **Social Media Aware Phrase Detection (SMAPD)** comments observed in social media interactions are scanned for terms or adjectives having no entry in the SentiWordNet set. If term or phrase is detected then it is given a polarity either positive or negative.

A list is prepared by searching Social media specific terms and phrases. This list is called "Social Media Aware Phrase List". Two different sources are used for the preparation of SMAPL (Social Media Aware Phrase List) (1) Existing data (2) Flagged comments. Those flagged comments are selected which have sentiments with high intensity. There are four major categories of comments: (1) **Self-promotion:** These are the comments which are used for subscription or for watching videos. (2) **Propaganda:** These are the comments which have strong beliefs on topics like Religion and socialism. (3) **Abusive comments:** These are those comments which contain extremely racist comments. (4) **Miscellaneous comments:** These are the comments which cannot be classified into any categories because the content of these comments looks normal.

Term sentiment identification is very difficult in a sentence. Only terms in a sentence are not sufficient for the identification of sentiment. For example, the word "pretty" having a positive sense is when paired with qualifier "not" i.e. "not pretty" the polarity of the phrase changes from positive to negative sense. In order to detect sentiments properly, it is very important to identify such contextual situation [1, 20].

Comments in Social media are scanned to detect terms or adjectives but these comments are not present [21] in SentiWordNet set. This set can further be improved and used for comment classification. A text with four terms is scanned weather these terms are present or absent in the list of SentiWordNet set. If these terms or phrases are detected then they are assigned polarity values i.e. 0, 1 or -1.

### 2.3 Predicting YouTube Comments

YouTube is a website for sharing of videos. Through YOUTUBE videos can be uploaded by the users. The users can also view, and share videos .YOUTUBE uses technologies of Adobe Flash Video and HTML5 in order to display a wide variety of user-generated videos [18]. YOUTUBE includes educational, music and short clips videos. Most videos allow users to leave comments.

There are four types of YouTube comments [22] **(1) Short syllable comments:** These comments are usually positive but still retarded. **(2) Advertisements of any kind:** These comments are for the advertisement of any organization or company **(3) Negative criticism:** These comments are related to insulting someone. **(4) The insane, rambling argument:** These are comments on religious and political videos, and always against the ideologies support in the video.

These user comments are analyzed in order to predict polarity [19,24,25]. This aims at analyzing comments made on videos hosted on YouTube and predicting the ratings that users give to these comments. The ratings are basically number of people likes (positive rating) or dislikes (negative rating) [4, 23, 24]. The authors refer to comments that have positive rating as accepted comments and those having negative ratings as unaccepted comments. When ratings of comments for videos are analyzed then it can provide indicators. By using these indicators high polarity of content can be achieved [4, 19, 26].

Using comment clustering and aggregation techniques, users of the system are provided with different views on that content. Further, different classifiers are built in order to estimate the ratings of comments. Through those classifiers comments can be structure and filtered automatically [4, 26].

To study online radicalization among YOUTUBE users, Bermingham et al. [26] analyzed user comments and the language which the users use for comments. In order to perform a high quality sentiment analysis, different techniques are proposed based on the method they use to select comments/words or sentences from YOUTU.

Figure 2 and figure 3 show some of the YOUTUBE comments:

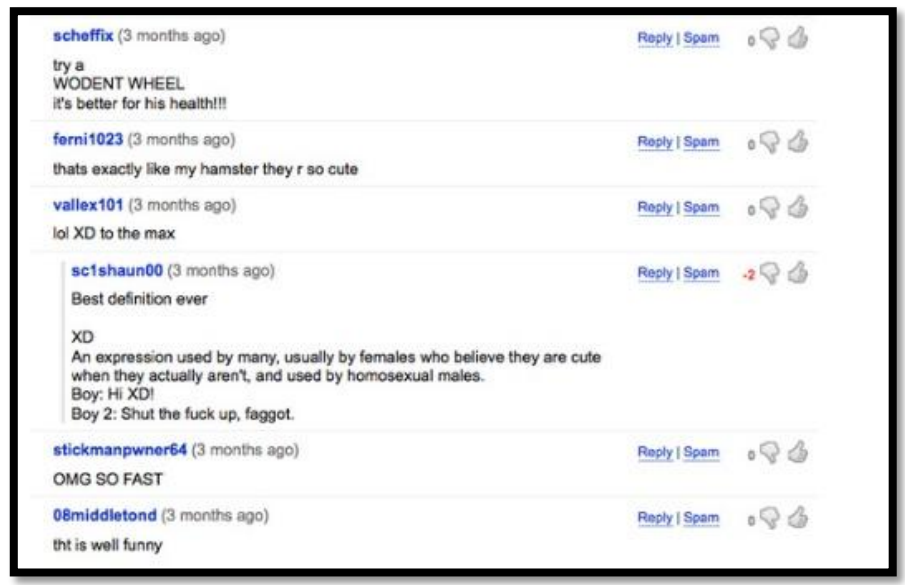

Fig. 2. Comments and Comment Ratings in YouTube

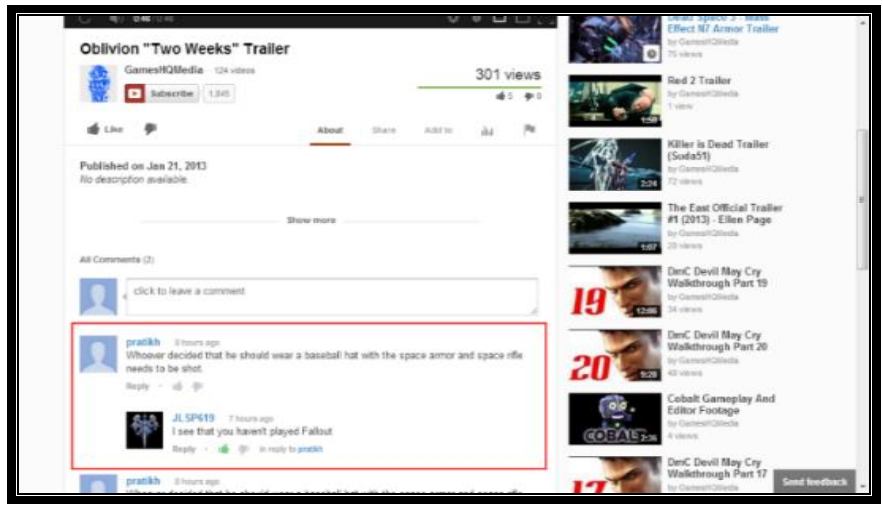

Fig. 3. Comments in YouTube

### 3. Evaluation

Precision and Recall are usually used to measure the accuracy of a Sentiment Analysis system. Precision or positive predictive value is a part of relevant retrieved instances [18], while recall or sensitivity is the fraction of retrieved relevant instances. So precision and recall are based on two facts, namely, relevant understanding and measure [27, 28].

$$\text{precision} = \frac{|\{\text{relevant documents}\} \cap \{\text{retrieved documents}\}|}{|\{\text{retrieved documents}\}|}$$

$$\text{recall} = \frac{|\{\text{relevant documents}\} \cap \{\text{retrieved documents}\}|}{|\{\text{relevant documents}\}|}$$

Table 2 shows the detailed results for each category.

For example, a video has to be analyzed containing 11 dogs and some cats. The program which is used for recognizing dogs in that video identifies only 7 dogs. If 4 dogs out of 7 identified dogs are identified correctly and remaining 3 in 7 identified dogs are actually cats, then the precision of the program is 4/7 and its recall is 4/11.

Similarly if a search engine returns 20 pages and only 10 pages out of 20 are relevant and search engine fails to return 30 additional relevant pages then its precision is 10/20 = 1/2 and its recall is 10/40 =1/4

Classification is performed using support vector machine (SVM) [29, 30, 31]. Then video comments are divided into two parts. One part is called test data, and other part as training data. Test data changes after each test. Each video is described by a vector. After test, SVM classifier is used to classify category for each test vector.

Table 2. Results of the SVM-classifier [Adopted from Peter Schultes at el.]

| | News& Politics | Comedy | Shows | Sports | Science & Technology | Gaming | People &Blogs | Films &Music | Animation | Entertainment | Pets & Animals |
|---|---|---|---|---|---|---|---|---|---|---|---|
| Precision | 1.00 | 0.88 | 1.00 | 1.00 | 1.00 | 1.00 | 1.00 | 1.00 | 1.00 | 1.00 | 1.00 |
| Recall | 1.00 | 1.00 | 1.00 | 1.00 | 0.67 | 0.67 | 0.89 | 1.00 | 1.00 | 1.00 | 1.00 |
| F-Score | 1.00 | 0.93 | 1.00 | 1.00 | 0.80 | 0.80 | 0.94 | 1.00 | 1.00 | 1.00 | 1.00 |

## 4. Conclusion

Classification of general events and detection of Sentiment Polarity of user comments in YouTube is a challenging task for researchers so far. A lot of work is done in this regard but still have a long way to go to overcome this problem.

In this paper we have emphasized on following problems in order to find the polarity of comments given by the users of YOUTUBE.1) Current sentiment dictionaries having limitations.2) Informal language styles used by users, 3) Estimation of sentiments for community-created terms, 4) To assign proper labels to events, 5) Achieve satisfactory classification performance 6) Challenges involving social media sentiment analysis. Different techniques like User Sentiment Detection, Event Classification and Predicting YOUTUBE comments used for comments polarity are also discussed.

Regarding future work, improving the social lexicon and to validate it statistically and proper event classification can help to increase the performance to predict rating of comments. Table 3 gives the comparative study of sentiment analysis on YOUTUBE.

**Table 3. Comparative Study of sentiment Analysis on YOUTUBE**

| Sr. No. | Methodology | Technique(s) | Tools Used | Dataset | Outcome | Study |
|---------|-------------|--------------|------------|---------|---------|-------|
| 1 | Un-Supervised Lexicon-based Sentiment Approach | Content preprocessing | YouTube API | YouTube Videos | Detected User | 1 |
| 2 | A Breadth-first Search | Clustering | YouTube API | Data collected in 3-month period from YouTube | Clustering of Videos | 2 |
| 3 | SentiWordNet Thesaurus | SentiWordNet-based Analysis of Terms | Google's Zeitgeist archive, queries | 67,000 YouTube Videos | Filter New Unrated Comments | 3 |
| 4 | Setiment Analysis | SentiWordNet/ Preprocessing | Lexicon Based Approach | Movie Reviews | Polarity of Text | 4 |
| 5 | Knowledge-based System | WordNet, SentiWordNet | Antelope NLP Framework | News Headlines | High Accuracy on Emotions and Valence Annotation | 7 |
| 6 | Opinion Mining | Bayesian Logistic Regression Classifier | Lemur Toolkit based on Language Modeling | News Articles Product & Movie Reviews | Track Opinion Mining Task | 12 |
| 7 | A Lexical Cohesion Based Sentiment Intensity & Polarity in Text | Precision, Recall, F-score | Cohesion Based Text Representation Algorithm | Financial News | Polarity Measurement in Text | 20 |
| 8 | Fusion Method | Sentiment Classification | Stanford Tagger | Opinion Reviews | Opinion Classification | 25 |
| 9 | Centroid Classifier | Vector Space Model | Similarity & Relative Similarity Ranking | Computer, Education , House Reviews | Sentiment Classifier on New Domain | 26 |
| 10 | Un-supervised Lexicon based Approach | Sentiment Analysis | YouTube Crawl | A Group Within YouTube | Gender Differences of Users, Suggesting Views of Female Users | 27 |